# Mini Volume Collapse as Evidence for a 3-Body Magnetic Polaron in Sm$_{1-x}$Eu$_x$S

Daniel Schaller[1], Patrick G. LaBarre[1], Tiglet Besara[2,4], Ernst Bucher[5], Theo Siegrist[2], and Arthur P. Ramirez[1]

[1]*Physics Department, University of California Santa Cruz, Santa Cruz, CA 95064, USA.*
[2]*FAMU-FSU College of Engineering, Tallahassee, FL 32310, USA.*
[3]*National High Magnetic Field Laboratory, Tallahassee, FL 32310, USA.*
[4]*Department of Physics, Astronomy, and Materials Science, Missouri State University, Springfield, MO 65897, USA*
[5]*International Solar Energy Research Center Konstanz, 78467 Konstanz, Germany*

**Abstract**

Samarium sulfide (SmS) is a non-magnetic narrow-gap (0.06 eV) semiconductor which undergoes a transition to a metallic intermediate valence state at 6.5 kbar. Europium sulfide (EuS) is a ferromagnetic semiconductor with a Curie temperature of 16K and a gap of 1.6 eV. Here we present a study of the lattice constant, magnetic susceptibility, and resistivity of the substitution series Sm$_{1-x}$Eu$_x$S for $0 < x < 1$. We observe a smooth interpolation of magnetic and transport behavior across the series, consistent with a virtual crystal scenario and Vegard's law. Surprisingly, however, the lattice constant deviates below Vegard's law in a manner that suggests parametric control of the Sm-Sm distance by the Eu moment in the manner of a magnetic polaron.

Interest in narrow band gap (NBG) semiconductors has grown with the advent of topological insulators (TIs), a key ingredient of which is a band gap energy less than the spin orbit coupling (SOC) strength, the largest values of which are in the range 0.6 eV for $5d$ elements. Among potential TI materials are the *Kondo* Insulators (KIs), whose gap results from hybridization of *f*-electrons with a conduction band [1]. Indeed such a topological Kondo insulating state was predicted by Dzero et al. in the KI material $SmB_6$ [2] and evidence for a surface conduction channel distinct from the bulk was found in non-local transport measurements [3, 4]. Another KI that has been studied intensively for its intermediate valence state is the chalcogenide NBG semiconductor SmS [5]. With a rock-salt structure, SmS possesses a band gap that varies among samples but is generally in the range of 0.06 eV [6]. Early work established the existence of a pressure-driven phase transition at 6.5 kbar from the black semiconducting phase with a magnetic singlet ground state to a metallic "golden" phase exhibiting magnetic behavior consistent with a Sm valence state of 2.8+ [5-15]. This metal-insulator transition has come under renewed scrutiny with a prediction of a concomitant change in topological invariant, with the golden phase seen as the metallic surface state of a topologically non-trivial insulator [13]. This prediction has been questioned, however, in a dynamical mean field theory (DMFT) study [14] that includes the effects of temperature-dependent hybridization among the Sm $4f_{5/2}$ and $5d$ ($t_{2g}$) electrons, effectively reproducing the major features of the resistivity vs. temperature of the golden phase [12]. Instead of the anticipated Dirac-cone dispersion, however the DMFT study showed that the surface states possess a small gap, suggesting that they are spin-polarized Rashba states. Intriguingly, DMFT shows the presence of two Fermi surface topology changes, at $T = 37K$ and 13K, related to the formation of the $4f_{5/2}$ band and a splitting within this band, respectively. Thus, while these results show that the SmS golden transition is not topological per se, the low energy of these temperature-driven Lifshitz transitions suggests that further Fermi-surface rearrangement might be possible by exploring neighboring phases.

Here, we present measurements of the lattice constants, the magnetic susceptibility ($\chi$), and the resistivity ($\rho$) of the solution series $Sm_{1-x}Eu_xS$ for $0 < x < 0.75$ at ambient pressure and temperatures down to 2K in order to assess the effect of introducing magnetism into the SmS system. Both SmS and EuS possess face centered cubic (fcc) rock salt structures with similar-size lattice constants (a=5.96 Å and a=5.97 Å respectively [16]) which will minimize chemical pressure effects. Such effects are observed when smaller rare-earth elements (*RE*) other than Eu are substituted into SmS to form $Sm_{1-x}RE_xS$. These systems undergo an explosive first order transition above the liquid nitrogen boiling temperature for x greater than a critical value [17]. Substitution of Eu produces no such effect over the entire range of x, which may be attributed to the similarity of lattice constants and the robustness of the fcc lattice [15]. In a previous study of $Sm_{1-x}Eu_xS$, the pressure dependence of $\rho$ at room temperature was used to explore the vanishing of the metal-insulator transition [16]. Here we study the behavior of $\rho(T)$ and $\chi(T)$ at ambient-pressure over a wide range of *T*. Whereas SmS is non-magnetic, exhibiting a van Vleck susceptibility below 100K, EuS is one of the few insulating ferromagnets, with a Curie temperature $T_c$ = 16.5K. The semiconducting gap of EuS is 1.6 eV [18], thus substantially larger than the gap in SmS and one might expect an interplay of magnetism and transport at the boundaries between Sm-rich and Eu-rich regions.

Single crystal samples of $Sm_{1-x}Eu_xS$ were prepared in a manner similar to synthesis of pure SmS, as described earlier [19]. High-resolution single crystal X-ray diffraction (XRD) measurements using a modified Bond method [20] provided the lattice constant change across the series. Magnetic susceptibility measurements, both dc ($\chi_{dc}$) as well as the real part of the ac-susceptibility ($\chi'_{ac}$), were made with a Quantum Design (QD) superconducting quantum interference device (SQUID) magnetometer. The $\chi_{dc}$ data were obtained with *H* = 0.1T and the $\chi'_{ac}$ measurements were obtained at *H* = 0 and at a frequency of 10 Hz. Resistivity measurements were made using the four wire method in a QD Physical Property Measurement System (PPMS). To minimize contact resistance, gold pads were evaporated onto the surface of the crystals, to which gold wires were attached via silver epoxy (Epo-tek H20E).

Seven concentrations of $Sm_{1-x}Eu_xS$ were studied, including pure SmS and EuS. In Fig. 1, the lattice constants (*a*) of $Sm_{1-x}Eu_xS$ are shown. The 0.01 Å decrease in *a* on going from x = 0 to 1, reported in previous work [16], is not observed in the present measurements, where *a* = 5.955 $\pm$ .001 Å and 5.957 $\pm$ .001 Å for x = 0 and 1 respectively. A substantial decrease in *a* by 0.015 Å is observed, however, for x = 0.10 and 0.25. This decrease is much larger than the standard deviation for the measurement and we will discuss a possible origin for this behavior below. It is important to mention that the lattice parameter decrease at the insulator to metal transition at 6.5 kbar in SmS is approximately 0.28 Å [21], much larger than the changes observed here on alloying with Eu.

In Fig. 2 are shown $\chi(T)$ and $\chi^{-1}(T)$ across the dilution series. For x = 0, the expected low-temperature van Vleck susceptibility ($\chi_{VV}$) is observed and for $0.05 \leq x \leq 0.25$, Eu substitution results in an additional paramagnetic contribution to $\chi(T)$. It is of interest to ask if the Eu moment is the value expected for the divalent $4f_{7/2}$ ion. Our knowledge of the Eu concentration, x, comes from the proportions of Sm and Eu in the starting material. We can test this by *assuming* that the paramagnetism is given by $Eu^{2+}$, which has an $S_{7/2}$ configuration and an expected moment of 7.94 $\mu_B$. In Fig. 3 (upper inset) are shown $\chi^{-1}(T)$ for x = 0.05, 0.10, and 0.25, with $\chi_{VV}$ subtracted. For this subtraction, we can use $\chi_{VV}$ for SmS, even though magnetic impurity ions will modify $\chi_{VV}$ since the sum of all three exchange interactions is the virtually the same for $Sm^{2+}$ as for $Eu^{2+}$ [22]. From these data we derived effective x-values of 0.067, 0.11, and 0.27 for these samples respectively which, given the uncertainty in the $\chi_{VV}$ subtraction and the neglect of clustering effects, is consistent with the nominal Eu concentration values. In the lower inset of Fig. 3 are shown ac-susceptibility ($\chi_{ac}$) data for the x = 0.25, 0.50, 0.75, and 1.0 samples close to their Curie temperatures, $T_c$. For x = 0.75 and 1.0 and below $T_c$ the data are demagnetization-limited and have been expanded for clarity. The finite, albeit small positive slope of $\chi(T)$ for these concentrations is most likely related to variation of the demagnetization field across the sample and related domain wall motion. The variation of $T_c$ with x is shown in Fig. 3. While our data do not extend below $T$ = 2K, they are consistent with a $T_c$ going to zero close to the site percolation threshold, x = 0.20, for the fcc lattice and short range interactions. Finally, we found no evidence of spin glass behavior in any of the compounds, also consistent with simple percolation physics.

In Fig. 4 are shown $\rho(T)$ data on $Sm_{1-x}Eu_xS$ for various x. Here, the low-temperature limit over which data are shown is that below which the resistance of our samples exceeded the upper limit of the resistance bridge. Accordingly, data for EuS were not obtainable. One sees that the main effect of substituting Eu for Sm is an increase in the magnitude of $\rho(T)$ over the measurement range. In Fig. 5 are shown the same data, but plotted as $\ln \rho$ vs. $1/T$. The behavior across the series is well-described by $\rho(T) = \rho_0 e^{(\Delta/2)/k_B T}$, as expected for an intrinsic semiconductor, with gaps given by $\Delta$. These data show that the origin of the increase in magnitude of $\rho$ is an increased gap for charge transport. Our measurements of pure SmS yield $\Delta/2$ = 426K, or 36meV, which corresponds to $\Delta$ = 72meV, close to the value reported by optical absorption experiments of 60meV [23]. The other concentrations and band gap energies are reported in table 1.

The systematic variation of $\Delta$ with x is shown in Fig. 6. We see that the gap smoothly interpolates between the values of the two pure systems, x = 0, 1. Such behavior is consistent with the virtual crystal approximation, often used in band structure calculations for systems with similarly mixed atomic constituents [24]. Also shown in Fig. 6 is the variation of $\theta_W$, extracted from fits of $\chi^{-1}(T)$. Here we depict two separate regimes. For $x > 0.5$, behavior associated with a mean field variation where the number of nearest neighbors is proportional to $x$ is observed. For $x < 0.5$, which is below the nearest neighbor percolation threshold (~0.2 for a simple cubic lattice) a finite $\theta_W$, associated with long range dipole interactions, is observed.

We now return to the lattice constant variation with x. As shown in Fig. 1, unlike $\chi(T)$ and $\rho(T)$, $a$ does not linearly interpolate between the end member values, which are very similar in size, differing by $1.7 \times 10^{-4}$. Instead, $a$ shows a minimum on the Sm-rich side of the series. The similarity in magnitude of $a$ between SmS and EuS was noted early on [15] and it makes the roughly 0.33% deviation in $a$ (1% decrease in volume) for $x < 0.5$ all the more striking. Such a volume collapse would also be caused by an applied pressure of 1.5 kbar, given that the bulk modulus is B = 151 kbar [21]. Since $\partial \chi / \partial P = 7.5 \times 10^{-5}$ emu/mole-kbar [5] at room temperature, the volume collapse implies a change in $\chi_{VV}$ of $1.1 \times 10^{-4}$ emu/mole, a 2.3% increase over the ambient pressure $\chi_{VV}$ of SmS. Such an increase is not observable directly in our $\chi(T)$ measurements due to the the overwhelming influence of the paramagnetic Eu$^{2+}$ moments, and is not large enough to affect the previous analysis of the effective concentration of these moments. Even though the overall effect on $\chi$ is only 2.3%, the volume collapse occurs at concentrations as low as $x = 0.1$, which implies that the increase of Sm $\chi_{VV}$ is much larger for those ions closest to the dilute Eu$^{2+}$ ions. We propose that this local enhancement of the Sm $\chi_{VV}$ is due to the large S = 7/2 moment of Eu$^{2+}$. Such a local field would have the effect of splitting the $J = 1$ excited multiplet, thus lowering the $J_z = -1$ state (where z is the direction of the nearby S = 7/2 moment), and increasing the susceptibility via $\chi_{VV} = 8N\mu_B^2/(\Delta + 8\sum_i Z_i J_i)$ where $Z_i$ is the number of $i$th equivalent nearest neighbors and $J_i$ is the antiferromagnetic exchange interaction at the $i$th distance [22]. Then, because $\partial \chi / \partial P > 0$, such an increase in $\chi$ would lead to a local volume collapse around the Eu$^{2+}$ ion.

A normal magnetic polaron effect as just described would produce a volume collapse proportional to the product of Eu and Sm concentrations, i.e. $a \propto x(1-x)$. Such a function has an extremum at $x = 1/2$, whereas we clearly see the minimum in $a(x)$ for $x \approx 0.3$. This suggests that the Eu$^{2+}$ ion is acting on the Sm-Sm bond instead of an isolated Sm ion. We propose that the controlling configuration is for a Eu$^{2+}$ ion simultaneously splitting the excited states of two Sm neighbors, leading to a large local contraction. In this case, the size of the contraction varies both in proportion to the Eu concentration *and* in proportion to the square of the Sm concentration. In other words $a = a_V - \mathcal{A} n_{Eu} n_{Sm}^2$, where $a_V$ is the Vegard's law interpolation between $a_{Sm}$ and $a_{Eu}$, $\mathcal{A}$ is a constant, and $n_{Eu}$ and $n_{Sm}$ are the densities of Eu and Sm respectively. The constant $\mathcal{A}$ would be obtained from an effective medium calculation that considered all of the possible nearest Sm neighbors to the $Eu^{2+}$ impurity. Here we treat this as a parameter and fit the lattice constant data to $a(x) = a_V - \mathcal{A} x(1-x)^2$, the result of which is shown in Fig. 1. We see that our assumptions represent an adequate description of the data and yield a value of $\mathcal{A} = 0.13$Å.

It is useful to rationalize the novel result of the non-Vegard's law behavior with the monotonic behavior of $\chi(T)$ and $\rho(T)$ as a function of $x$. For $\chi(T)$, we see essentially single-spin behavior for $x$ less than the percolation threshold and interacting spin behavior for $x$ above this value. For $\rho(T)$, we find a gap that smoothly interpolates between the gaps of SmS and EuS. This is consistent with a virtual crystal approximation, given a mean free path for electron transport much greater than the inter-atomic distance. For both quantities, our interpretation of non-Vegard's law behavior in $a(x)$ will not significantly affect these interpretations at the present level of experimental precision.

Acknowledgments: We acknowledge useful discussions with B. I. Shklovskii and S. von Molnar. The low temperature measurements were performed at UCSC and the Xray diffraction measurements were performed at FSU/NHMFL. A. P. Ramirez and P. G. LaBarre were supported by U.S. Department of Energy Grant No. DE-SC0017862. T. Besara and T. Siegrist performed the lattice constant measurements and acknowledge support from the National Science Foundation, Division of Materials Research, NSF-DMREF 1534818. Part of the work was carried out at the National High Magnetic Field Laboratory, which is supported by the National Science Foundation under grant NSF DMR 1644779, and by the State of Florida.

Table 1

| x (Eu) | $T_c$ (K) | $\mu_{eff}$ (μ$_B$) | $\Delta/k_B$ (K) |
|---|---|---|---|
| 0 | -- | -- | 426 ± 1 |
| 0.05 | -- | -- | 620 ± 11 |
| 0.1 | -- | -- | 735 ± 2 |
| 0.25 | 3.4 ± 0.1 | 8.3 ± 0.2 | 1380 ± 37 |
| 0.5 | 9.9 ± 0.1 | 7.9 ± 0.2 | 2370 ± 62 |
| 0.75 | 13.9 ± 0.1 | 8.1 ± 0.1 | 2962 ± 152 |
| 1 | 16.5 ± 0.1 | 8.1 ± 0.1 | -- |

Table 1: Transition temperature to the ferromagnetic state, effective moment ($\mu_{eff}$), and band gap energy (Δ) for various concentrations of $Sm_{1-x}Eu_xS$.

Figures:

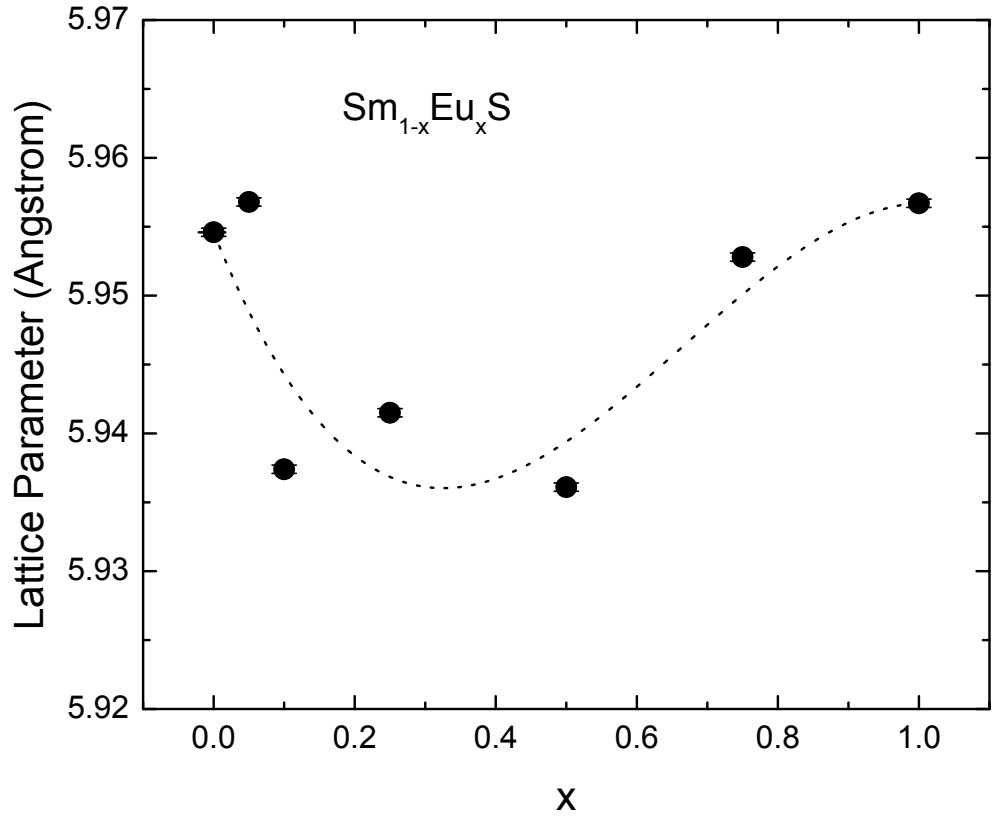

Figure 1: Lattice constants of $Sm_{1-x}Eu_xS$ at 300K vs concentration x. The dotted line represents a lattice constant that varies as $a(x) = (a_{Sm}n_{Sm} + a_{Eu}n_{Eu}) - \mathcal{A}n_{Sm}^2 n_{Eu}$, as discussed in the text.

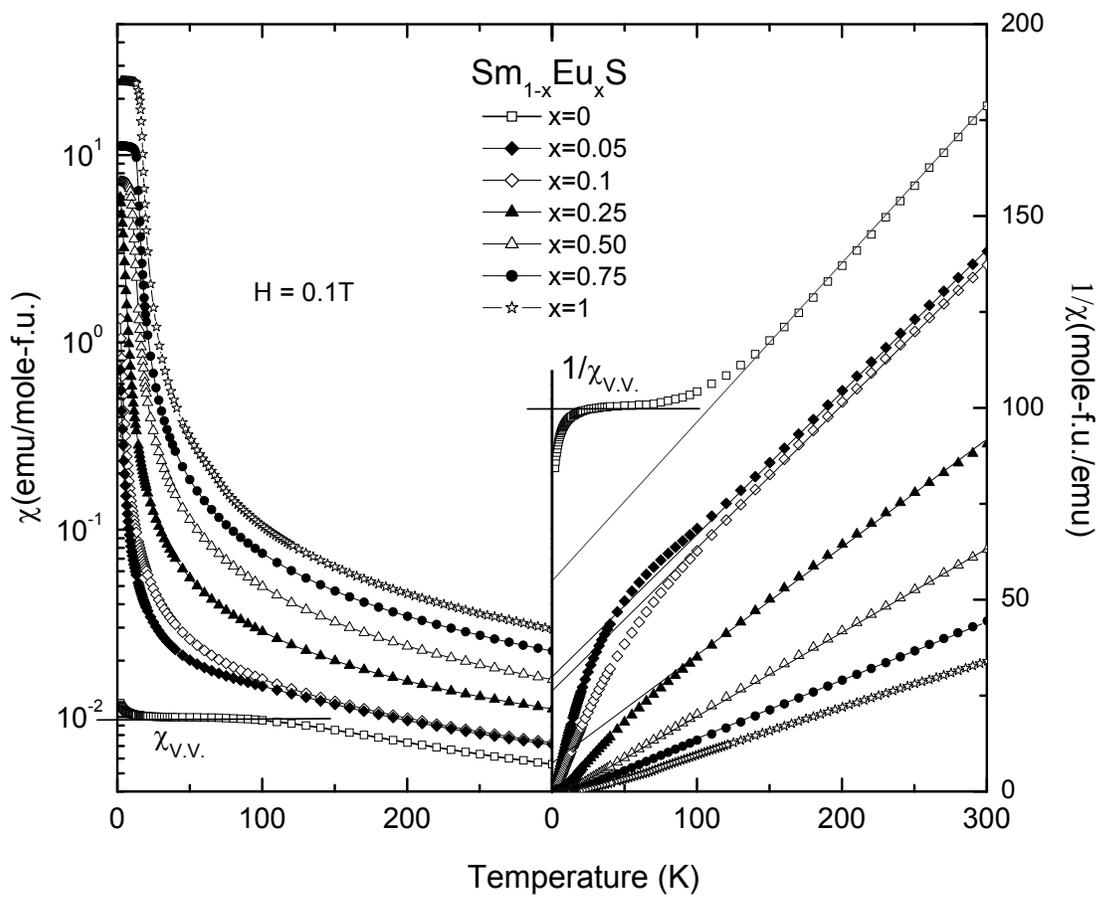

Figure 2: Magnetic susceptibility and inverse susceptibility of $Sm_{1-x}Eu_xS$.

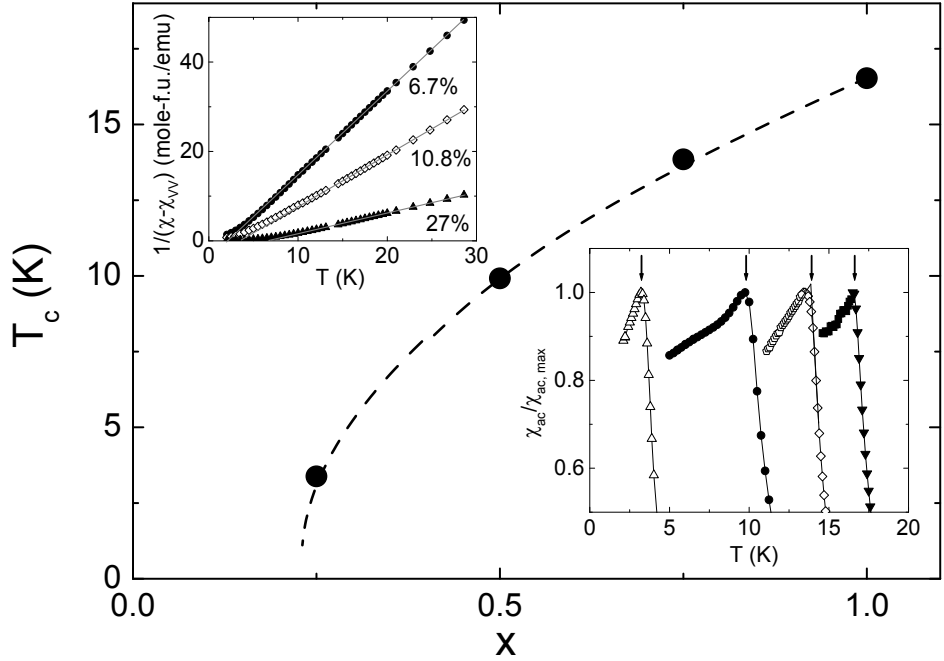

Figure 3: Upper inset: Percent composition of Eu in $Sm_{1-x}Eu_xS$ based on the expected $Eu^{2+}$ magnetic moment. Lower inset: The ac-susceptibility of $Sm_{1-x}Eu_xS$ for x = 0.25, 0.5, 0.75, and 1 (left to right) divided by the peak value at temperatures close to the ferromagnetic $T_c$, indicated by arrows and given in Table 1. The deviation of the data from unity for x = 0.75 and 1.0 below $T_c$ have been multiplied by 10 and 100 respectively to make the peak more apparent. Main figure: The values of $T_c$ versus x for $Sm_{1-x}Eu_xS$ for x = 0.25, 0.50, 0.75 and 1.0. The dashed line is a guide to the eye and suggests a percolation threshold of approximately 0.20.

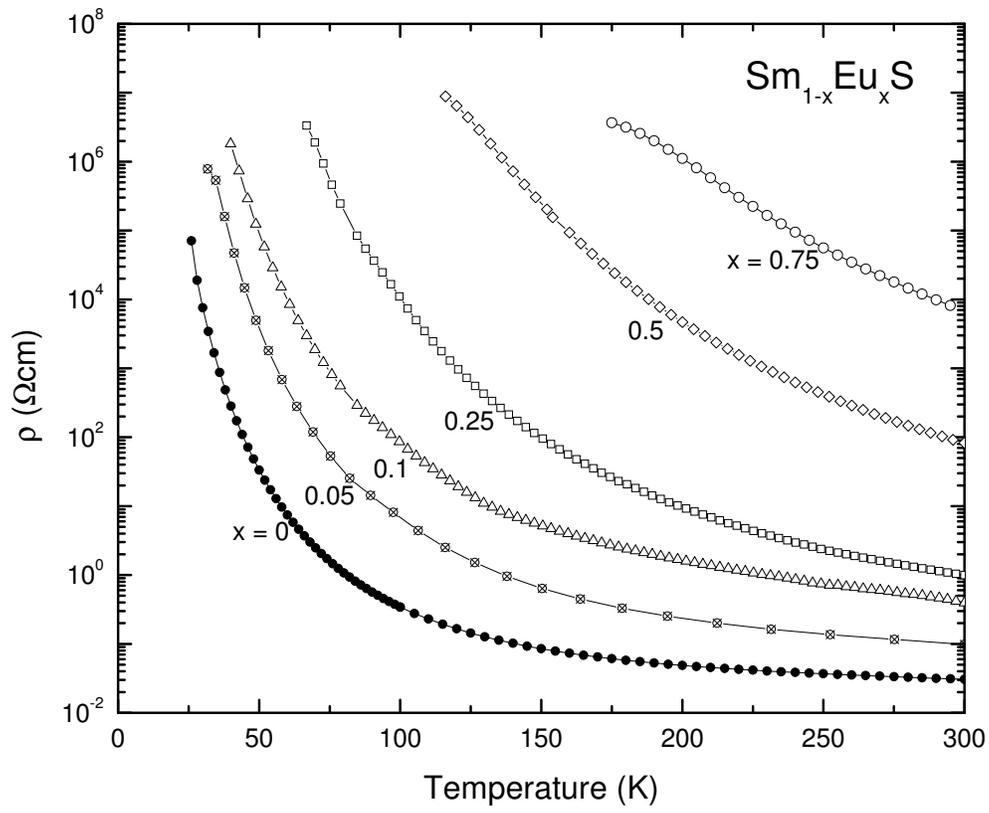

Figure 4: Resistivity of different concentrations of $Sm_{1-x}Eu_xS$. Data for x = 1 were not obtainable due to an upper limit on the measured resistance of the apparatus.

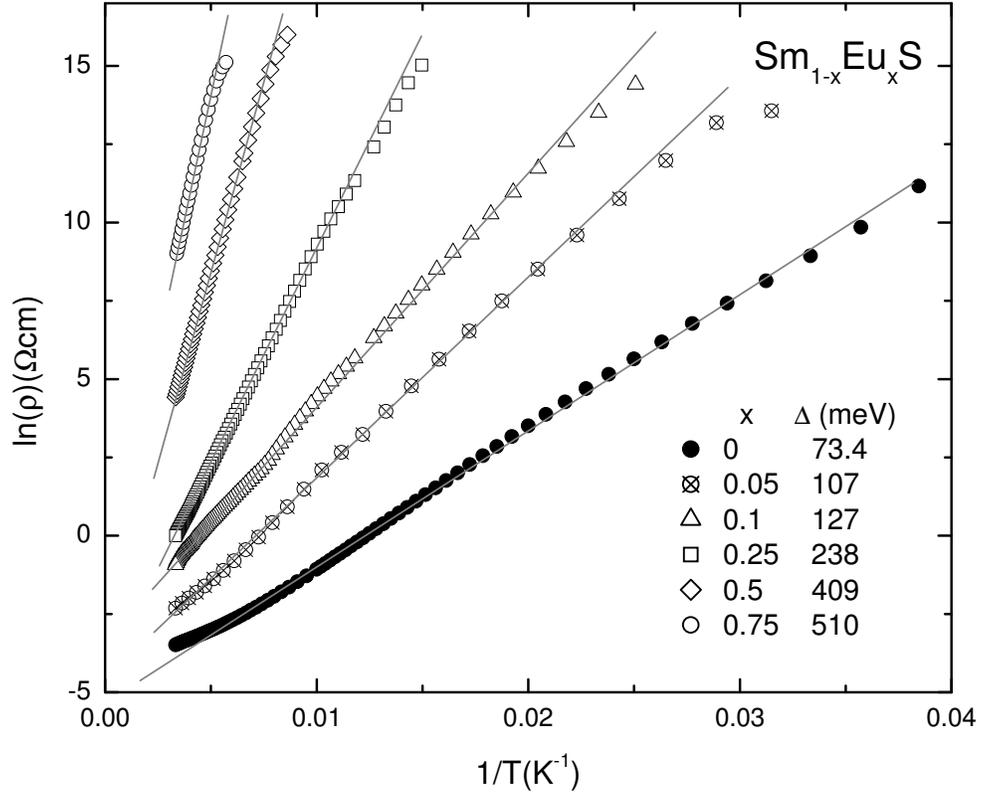

Figure 5: Log of resistivity versus inverse temperature for $Sm_{1-x}Eu_xS$. The straight lines represent fits to $\rho(T) = \rho_0 e^{(\Delta/2)/k_B T}$ and $\Delta$ values are given in Table 1.

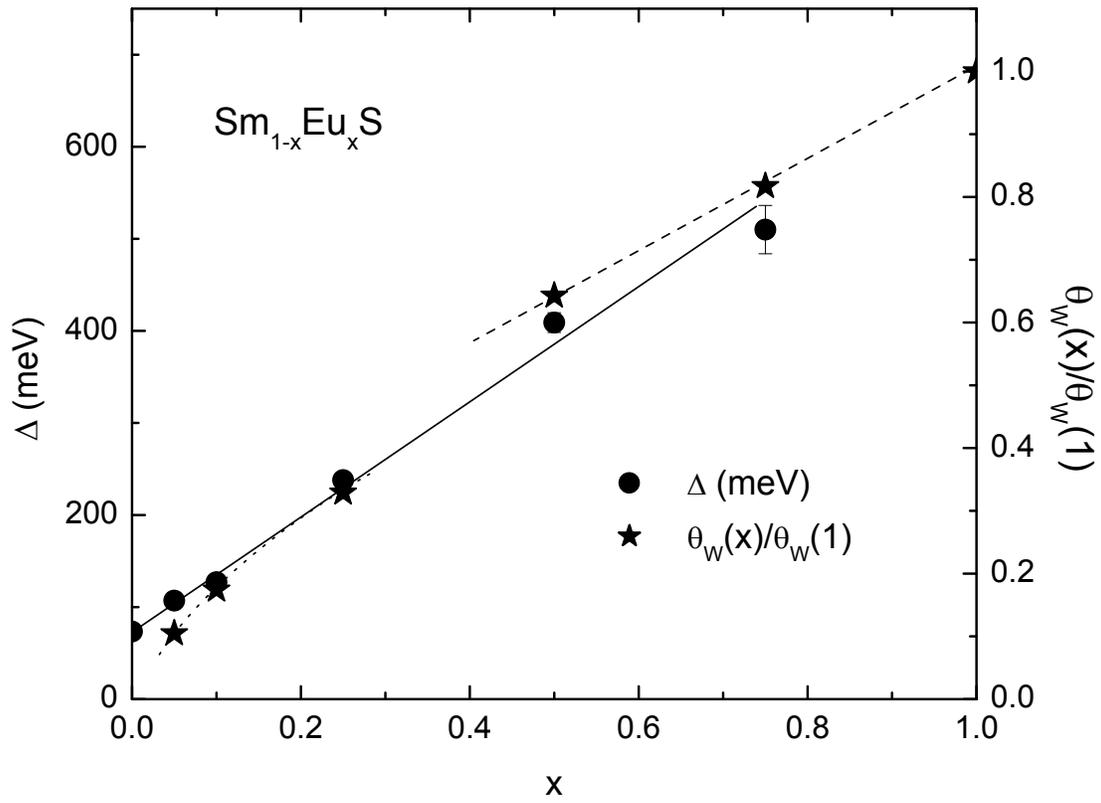

Figure 6: Circles: Transport gaps, Δ, as a function of percent composition in $Sm_{1-x}Eu_xS$. Stars: The Weiss constant, $\theta_W$, as a function of x.